# Coherent anti-Stokes Raman scattering through thick biological tissues by single wavefront shaping

**Matthias Hofer,**[1] **Siddarth Shivkumar,**[1] **Bilal El Waly,**[2] **and Sophie Brasselet**[1,*]

[1]*Aix Marseille Univ, CNRS, Centrale Marseille, Institut Fresnel, F-13013 Marseille, France*
[2]*Institut des Neurosciences de la Timone, Aix-Marseille Université and CNRS UMR7289, Marseille, France*

*\*sophie.brasselet@fresnel.fr*

**Abstract:** Coherent Anti Stokes Raman Scattering (CARS) offers many advantages for nonlinear bio-imaging, thanks to its sub-cellular spatial resolution and unique chemical specificity. Its working principle requires two incident pulsed laser beams with distinct frequencies to be focused in space and time, which focus quality however rapidly deteriorates when propagating at large depths in biological tissues. The depth limits of CARS and the capability of wavefront correction to overcome these limits are currently unknown. In this work we exploit the spectral correlation properties of the transmission matrix of a scattering medium in a pulsed regime, to recover coherent focusing for two distant incident CARS wavelengths which propagation is initially uncorrelated. Using wavefront shaping with a single spatial light modulator, we recover CARS generation through thick mice spinal cord tissues where initially no signal is measurable due to scattering, and demonstrate point scanning over large field of views of tens of micrometers.

## 1. Introduction

Nonlinear optical imaging is a key approach for deciphering biological processes in tissues with sub-cellular spatial resolution. Considerable progresses in this field have permitted to bring its application range to biomedical questions, in particular for the understanding of fundamental processes at the origin of pathologies, from cancer to neurological disorders [1,2]. Besides second and third harmonic generation, which are extensively used, Coherent Anti-Stokes Raman Scattering (CARS) is particularly interesting since it allows chemically specific label free imaging of targeted molecular vibration bonds. CARS is a third order nonlinear process that involves two near infra-red laser pulses of frequencies $\omega_p$ (pump) and $\omega_s$ (Stokes) in a resonant four wave mixing radiation at the emission frequency $(2\omega_p - \omega_s)$ (Fig. 1(a)). At resonance, the difference $\omega_p - \omega_s$ matches a specific vibration bond frequency. Typically, wavelengths differences ranging from 50 nm to 250 nm allow imaging wavenumbers from 800 $cm^{-1}$ to 3000 $cm^{-1}$, which covers essential chemical bonds present in biological samples. CARS has been applied to tissue lipid, proteins and chemical species imaging [2]. To perform CARS imaging in biological samples, stringent conditions need however to apply: both laser beams need to be overlapped in time and space, within tight foci of high peak power to allow the nonlinear interaction to occur. In biological tissues of thicknesses close to their scattering mean free path (Ls ~ 100 μm in the brain [3,4]), space and time overlap is already strongly compromised by optical aberrations. In this regime where ballistic light is however still present, adaptive optics can correct for aberrations and increase CARS signals by optimizing the incident wavefront using a deformable mirror [5]. This approach relies on a starting CARS signal, which is possible only at shallow depths, typically in white chicken muscle breasts at 100-200 μm thickness [5]. At larger depths of several Ls approaching the transport mean free path of the sample (Lt ~ 1 mm) [4], the propagation is scrambled by multiple scattering events and optical phases and directions are lost. Millimeter-depth CARS imaging has been so far achieved only by the use of a miniature objective lens, which however requires the physical insertion of a device in the tissue [6]. In this regime, light propagates as random field patterns, i.e. speckle, where the coherent superposition of the pump and Stokes CARS excitation beams is no more guaranteed. Nevertheless there is still a large amount of incident photons to benefit from, and several strategies have been developed to exploit scattered photons and recover a degree of spatio-temporal coherence after propagation through a scattering medium. Optimization schemes [7] or a direct measurement of the transmission matrix (TM) of the medium [8] have permitted to refocus light behind a scattering medium, opening a new field for imaging in scattering media. This tool has been successfully applied to second order nonlinear processes that use a single incident frequency, such as two photon fluorescence [9,10] and second harmonic generation (SHG) [11,12]. In CARS, the requirement to coherently manipulate two incident beams makes the situation more complex, and CARS generation behind a scattering medium a still unresolved challenge [2]. First, the presence of a few scatterers is already expected to strongly degrade the CARS signal due to a rapid modification of the nonlinear spatio-temporal coherent build-up [13]. Second, the chromatic nature of random waves' propagation in disordered media makes the pump and Stokes beams follow very different paths, resulting in uncorrelated speckles with a loss of mutual coherence.

Our proposed method is based on the use of a single wavefront correction for the coherent refocusing of both pump and Stokes excitation beams, in order to preserve their mutual coherence and generate a CARS signal even in conditions where the incident beams resemble initially random uncorrelated speckles. To demonstrate the potential of randomly propagated waves to recover mutual coherent refocusing and CARS generation by wavefront shaping, we monitor both linear and nonlinear optical properties at the exit of thick scattering media, i.e. in the transmission geometry. This approach is based on the manipulation of the Transmission Matrix (TM) of the scattering medium, which quantifies the intrinsic relation between input and output optical modes propagating through it [8]. The medium TM permits advantageously to create any wanted focus at the exit of the medium, and allows point scanning capabilities [12]. More importantly, the TM knowledge is particularly useful to evaluate spectral correlation properties that are at the basis of the coherent manipulation of two pulses of different frequencies. In what follows we demonstrate that under the short pulse / broadband conditions used, the refocus formed is stable over a sufficiently large spectral detuning, which leaves degrees of freedom for manipulating two distant frequencies even in thick samples. This points out the remarkable property of the broadband TM which, by construction, leads to enhanced angular [14], spectral [15] and polarization [16] memory ranges. We then develop different strategies for the construction of a new TM for CARS, which enables refocusing of both pump and Stokes beams. By the use of this new TM, we demonstrate the capacity of CARS signal generation through thick biological tissues at large wavelengths differences (~ 230 nm), typically used for the detection of lipids in the CH stretch spectral region (~ 2840-2950 cm$^{-1}$). We apply this approach to CARS generation through fixed slices of mice spinal cords of a few 100's μm thickness, a tissue that is known to be highly scattering [17] and where no initial CARS signal is measurable. At last, we show CARS point scanning over 10's of μm field of views, which is above the limits imposed by the angular memory effect of the medium and compatible with bio-imaging conditions. More generally the method is applicable for two input frequency mechanisms such as two-color two photon fluorescence and four wave mixing.

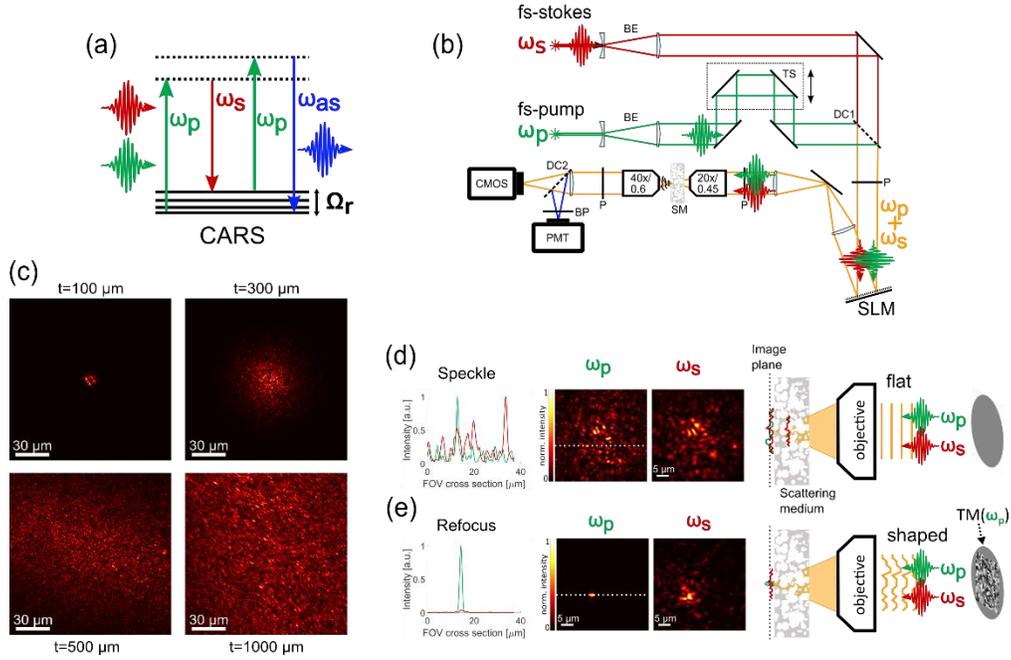

Fig. 1. CARS principle and wavefront shaping in scattering media. (a) CARS energy diagram depicting pump ($\omega_p$) and Stokes ($\omega_s$) frequencies interacting through a four wave mixing process to produce an anti-Stokes radiated frequency ($\omega_{as}$). (b) Set-up used for CARS generation through scattering media. BE: beam expander. DC1, DC2: dichroic mirrors. TS: translation stage. SM: scattering medium. SLM: spatial light modulator. BP: band pass filter. PMT: photomultiplier. sCMOS: camera. (c) $\omega_p$ image through different thicknesses (t) of spinal cord tissues. (d) Images obtained at both frequencies $\omega_p$ and $\omega_s$, without wavefront shaping (flat wavefronts), through a 300 µm thick spinal cord tissue. The red ($\omega_s$) and green ($\omega_p$) lines represent intensity profiles over the dashed line shown in the $\omega_p$ image. (e) Images obtained after refocusing using the TM measured for the $\omega_p$ frequency. Images profiles are depicted similarly as above.

## 2. Results

To generate CARS signals through scattering media, two focused short pulse beams coming from an optical parametric oscillator (pulse width at the laser output 200 fs, repetition rate 80 MHz, pump wavelength at 790 nm, Stokes at 1019 nm) were used. These wavelengths and short pulse widths allow CARS generation in the lipid CH stretch region at 2845 cm$^{-1}$, with a 150 cm$^{-1}$ spectral resolution. A delay line permits the time overlap of the two pulses at the exit of the sample plane, where the free-space focus is initially positioned. Both beams are reflected off a spatial light modulator (SLM) (Fig. 1(b)) used for wavefront manipulation. For refocusing the incident beams behind the scattering sample, the transmission matrix (TM) of the medium is measured as described in [8], at a position of 400 µm behind the scattering medium exit plane. The matrix relation between the incident and emerging field modes is extracted using self-reference step-interferometry where the SLM is used to phase-tune part of the incident wavefront on 1024

Hadamard basis components (32x32 macropixels on the SLM), while the other part serves as a reference beam (See Materials and Methods). Identical phase patterns are applied for both pump and Stokes beams, using a phase-to-voltage look up table measured at the mean frequency $\omega_m = (\omega_p+\omega_s)/2$ (see Section 1 of Supplement 1). As biological scattering samples, we used fixed mice spinal cord tissue slices (see Materials and Methods). Spinal cord is very rich in lipids, forming dense and tightly packed membranes around axons, called myelin sheaths, which have been extensively studied by CARS imaging at shallow depths below 100 µm [18,19]. Myelin's refractive index is higher than the rest of the cellular and extracellular matrix environment, which makes myelin-rich tissues particularly heterogeneous for optical propagation [17].

Fig. 1(c) shows typical images of the $\omega_p$ intensity patterns obtained after propagating the beam through fixed mice spinal cord tissue slices of increasing thicknesses. After a 100 µm thick slice, the focus starts being deteriorated, while at 300 µm depth a speckle is obtained, which is formed of fields randomly distributed in amplitude and phase over a large field of view. Since both pump and Stokes beams need to be jointly focused for CARS generation, the important parameter is not only the deterioration of the foci, but also their spatial overlap. While after 100 µm depth, a slight overlap between the transmitted pump and Stokes intensity patterns is still perceptible, this is not any more the case at 300 µm depth where the produced speckles are seen to be completely decorrelated (Fig. 1(d)). This decorrelation has a strong influence on the robustness of refocusing to spectral tuning. The refocus efficiency is quantified by its enhancement factor $\eta$, defined by the ratio between the refocus maximum intensity and the average speckle intensity with no wavefront correction. An example of refocussing by wavefront shaping is shown in Fig. 1(e), using TM($\omega_p$), the TM measured at $\omega_p$, to recover fields refocusing. At $\omega_p$, a maximum enhancement value of 360 is typically obtained with a 300 µm thick spinal cord tissue. Moving to the Stokes frequency $\omega_s$, the focused nature of the beam is lost and the enhancement drops to about 30, i.e. more than 10 times lower than the value obtained at $\omega_p$. Note that the chromatic dependence of the SLM contributes to only 4 % change in the obtained enhancements (see Section 1 of Supplement 1). The observed effect is therefore essentially due to spectral decorrelation of the measured TM in the medium. It means in particular that the strategy to form a refocus using wavefront shaping at $\omega_p$ will fail to form a refocus at $\omega_s$.

A way to quantify the spectral decorrelation mechanisms occurring in a scattering medium is to plot the progressive loss of correlation of the transmitted speckle when tuning the incident wavelength. The resulting spectral correlation dependence has been previously used to quantify the spectral width of a scattering medium [20]. It is also generally related to the spectral bandwidth of the medium, which gives the number of spectral modes that the medium contains with respect to the laser spectral width used for the illumination [21,22]. This dependence is shown in Fig. 2(a) for four different thicknesses of mouse spinal cord tissues from 100 μm to 1 mm, tuning the incident pump wavelength from 785 nm to 985 nm. Obviously, the correlation of produced speckles at the pump and Stokes wavelengths 790 nm and 1019 nm is expected to be very poor, even at a 100 μm thickness.

While speckle correlations carry information on how intensity patterns change when tuning the incident wavelength, the refocus formed by TM inversion is the result of a coherent interference that is performed through the sample propagation. Speckle and refocus spectral dependence can therefore differ depending on how the TM is measured. Ultimately for nonlinear scanning experiments, the important parameter is therefore the spectral memory of the refocus itself, which is quantified by the spectral dependence of the refocus enhancement factor as a function of the incident wavelength. This dependence is depicted in Fig. 2(b) for four different thicknesses of mouse spinal cord tissues from 100 μm to 1 mm, where the enhancements are normalized to their value at 790 nm. Remarkably, the spectral memory range of the refocus is larger than the spectral correlation width measured by speckle correlation. We attribute this to the fact that the TM manipulated in this experiment is measured under broadband conditions (i.e. the bandwidth of the incident lasers is similar or larger than the spectral bandwidth of the medium), as recently observed in thin scattering media [23]. Under such condition, it has been shown that the obtained refocus is formed essentially from short path photons, which survive the coherence conditions required to obtain the interferences necessary to measure the TM [15]. This effect has been shown to also lead to a remarkably large polarization memory range in the obtained refocus [16]. Here another consequence of this broadband TM is a large spectral memory range of this broadband refocus. Interestingly, these working conditions allow to reach a non-negligible degree of spectral correlations in spinal cord samples of a few 100's μm thickness for wavelengths differences of about 100-150nm, which correspond to typical vibration bond targeted in CARS experiments.

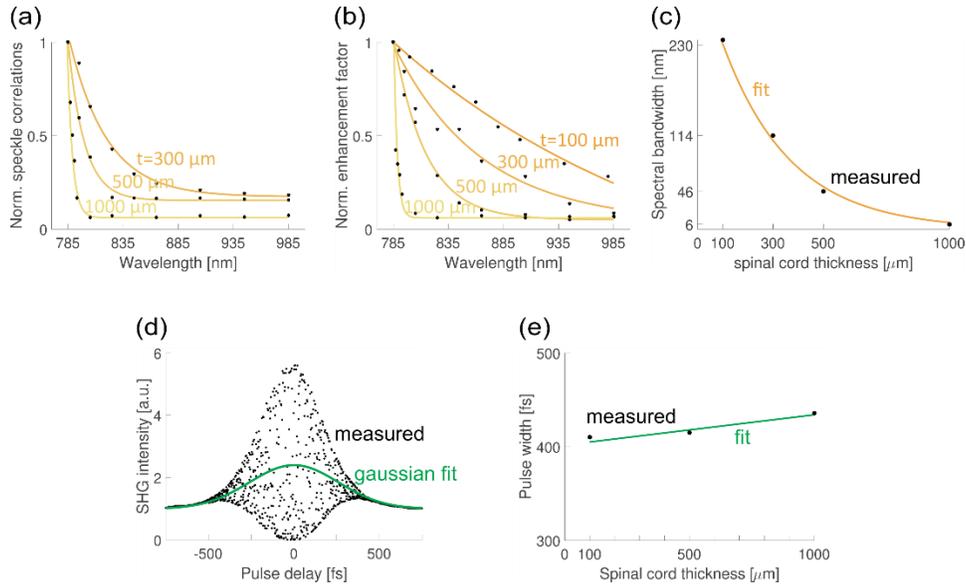

Fig. 2. Spectral and time properties of a refocus after wavefront shaping in spinal cord tissues. (a) Spectral decorrelation of the speckle pattern normalized to that obtained at 785 nm, as a function of the incident wavelength. (b) Dependence of the refocus intensity normalized to that obtained at 785 nm, as a function of the incident wavelength. The continuous lines in (a) and (b) are exponential fits (assuming symmetry around the central wavelength). (c) Full Width Half Maxima (FWHM) of the spectral dependence of b reported as a function of the sample thickness. The continuous line follows an inverse proportion dependence. (d) SHG autocorrelation measurement from a thin BBO crystal (200 µm), placed behind a 300 µm thick spinal cord slice. The SHG signal is obtained after refocusing by wavefront shaping. The fit of the pure autocorrelation contribution by a Gaussian shape is shown in green. (e) Measured pulse widths from foci obtained by wavefront shaping through different spinal cord thicknesses. The values reported are the FWHM of Gaussian fits from the SHG autocorrelation traces as shown in (c), i.e. the Gaussian width divided by $\sqrt{2}$. The line is shown as a visual guide.

When the spinal cord thickness increases above 500 µm however, the refocus spectral bandwidth decreases until reaching dramatically low values in very thick samples (e.g. 6 nm at 1000 µm thickness). Remarkably, the spectral width is roughly inversely proportional to the sample thickness (Fig. 2(c)), which is very different from what is expected in a multiply scattering regime where an inversely squared dependence is expected [22]. This emphasizes the specificity of both the broadband TM measurement method and the samples addressed, which, because their thickness does not surpass their transport mean free path, differ from a multiple scattering regime. Another important characteristic of the obtained refocus is its time width. Indeed, a requirement to ensure proper nonlinear coupling in a scattering medium is to preserve short pulses after wavefront shaping has been applied. To measure this pulse width, a second order autocorrelation measurement was performed by second harmonic generation (SHG) in a thin BBO crystal placed after the scattering medium. The pump beam (wavelength 790 nm) was split in two and a delay stage was used in one arm as a time-delay control. Both beams are reflected on the SLM and refocused into the BBO crystal, to reconstruct the typical autocorrelation SHG traces of Fig. 2(d) (see Section 2 of Supplement 1). The recorded time autocorrelation is fit by an envelope which assumes Gaussian shaped pulses, which width is reported in Fig. 2(e) as a function of the spinal cord sample thickness. Note that the initial pulse width measured at the sample position is 350 fs and therefore larger than the initial pulse width, which is attributed to dispersion from optical elements. Increasing the sample thicknesses from 100 μm to 1 mm is seen to increase the refocus pulse widths from 410 fs to 435 fs, which means that the pulse is not considerably broadened by the propagation through the scattering medium. This is another important consequence of the use of the broadband TM for the wavefront shaping correction: since the process of the TM measurement excludes long path photons in the sample, the obtained focus is not considerable enlarged in time [15]. This demonstrates that proper nonlinear coupling can operate at large widths in spinal cord tissues, as illustrated by the fact that a SHG signal can be obtained even at 1 mm thickness.

Refocus spectral and time measurements permit to draw limits for thicknesses to be used for wavefront shaping in scattering media at multiple wavelengths, in particular for CARS generation. The spectral memory curves shown in Fig. 2(b) show in particular, that a spinal cord sample thickness of 500 μm can only accommodate wavelengths differences less than 100 nm, which corresponds to wavenumbers below 1400 $cm^{-1}$. To reach the lipid bond region around 2845 $cm^{-1}$ as targeted here, a difference of about 230 nm is required. In this regime, the spectral memory of the refocus in fixed spinal cords is guaranteed only for very thin samples, below 100 μm thickness. Obviously even in thin media, applying a wavefront correction obtained at the pump wavelength for the Stokes beam will lead to very poor nonlinear CARS signal generation, since the Stokes beam will suffer from a very low refocus intensity (Fig. 1(e)). In what follows, we present strategies to compromise the use of both wavelengths in even thicker samples.

A first strategy is to use an incident wavefront that is obtained at an intermediate wavelength, benefiting from a shared spectral correlation extent by both pump and Stokes beams, with the use of the spectral memory of the refocus as depicted in Fig. 2(b). Fig. 3(a) shows the result of this strategy, where the TM of a 300 μm thick spinal cord sample is measured at $\omega_m = (\omega_p + \omega_s)/2$ (e.g. a wavelength of 905 nm for pump and Stokes respectively at 790 nm and 1019 nm). Using TM($\omega_m$) to refocus both pump and Stokes beams permits to obtain foci of similar shapes and sizes (Fig. 3(a)), with enhancements of 120 and 110 for $\omega_p$ and $\omega_s$ respectively. The enhancement value obtained at $\omega_s$ is about 4 times larger than that obtained with the inversion of TM($\omega_p$), as a consequence of a better spectral correlation between TM($\omega_s$) and TM($\omega_m$). An even more optimal situation for CARS generation is expected from a more appropriate balance between the pump and Stokes powers. Indeed CARS exhibits a nonlinear dependence proportional to the squared pump intensity, $I_{CARS} \propto I_p^2 \cdot I_s$ with $I_p$ and $I_s$ the pump and Stokes intensities, respectively. To adapt to this situation, the pump to Stokes power ratio has to be around 2, therefore a stronger weight has to be given to the pump wavelength in the wavelength chosen for TM measurement (see Section 3 of Supplement 1). This was obtained in practice by using, for the measurement of the TM, a wavelength of around 866 nm instead of 905 nm (the precise choice of the wavelength chosen varies upon the spectral memory and thus of the sample thickness, as detailed in Section 3 of Supplement 1). Using 866 nm as an intermediate wavelength, enhancements for both $\omega_p$ and $\omega_s$ were respectively 160 and 80.

The drawback of this strategy is that measuring the TM at a wavelength different from $\omega_p$ and $\omega_s$ requires to first tune the pump laser beam wavelength for the TM measurement, and then tune it back to $\omega_p$ to produce a CARS signal. To preserve measurement time and laser pointing stability, it might be of advantage to measure TMs independently at the respective pump and Stokes wavelengths prior to CARS generation. Using both TMs presents also the advantage to reconstitute a coherent addition of propagation conditions that is beneficial for both wavelengths in an optimized way with respect to phases. The coherent addition of TM($\omega_p$) and TM($\omega_s$) results in a composed TM$_{coh}$ that necessarily accommodates the refocus for $\omega_p$ and $\omega_s$ in a balanced way, even if not optimal for neither of them.

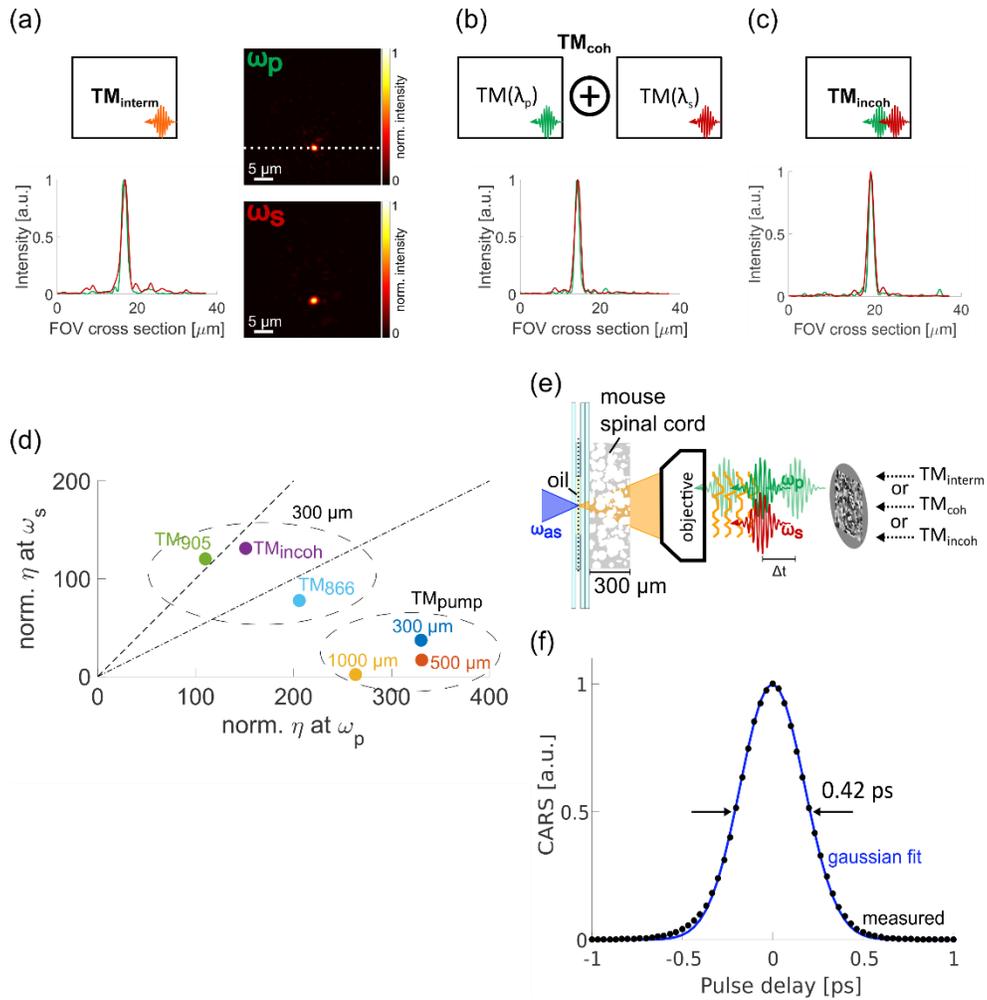

Fig. 3. Strategies for CARS generation behind a 300 μm thick mouse spinal cord tissue. (a) Use of the TM measured at an intermediate wavelength. The obtained refoci at $\omega_p$ and $\omega_s$ are depicted after refocusing with the use of $TM_{interm}$, as well as their lateral profile (here at 866 nm). (b) Use of a reconstituted TM, composed from the coherent addition of TMs measured at $\omega_p$ and $\omega_s$ ($TM_{coh}$). The profile of the obtained refoci at $\omega_p$ and $\omega_s$ are depicted. (c) Use of a TM measured from the incoherent addition of $\omega_p$ and $\omega_s$ beams ($TM_{incoh}$). The profile of the obtained refoci at $\omega_p$ and $\omega_s$ are depicted. (d) Enhancement values obtained for both $\omega_p$ and $\omega_s$ wavelengths, represented for situations described above, for different spinal cord thicknesses. The values obtained using $TM(\omega_p)$ are also shown (represented as $TM_{pump}$). $TM_{interm}$ is measured at both medium wavelength 905 nm and at 866 nm. The upper dashed line represents an equal ratio of pump and Stokes enhancement factors. The lower dashed line represents a ratio of 2, which is optimal for CARS generation. $TM_{coh}$ gives similar results as for $TM_{interm}$. (e) Schematics of the experiment used for CARS generation in an oil sample placed after the scattering medium. (f) Obtained CARS signal in the $TM_{interm}$ case, as a function of the delay time between the $\omega_p$ and $\omega_s$ beams. The signal is normalized to its value at zero delay.

To build this composed $TM_{coh}$, we simply add the contribution of the two $\omega_p$ and $\omega_s$ complex fields with equal weight. Because experiments were conducted with a phase-only SLM, we ignore the amplitude part of the TMs and only keep their phase contributions. This strategy permits to reach typical enhancement factors similar to the intermediate TM of 120 for both $\omega_p$ and $\omega_s$ refoci (Fig. 3(b)). For maximum CARS generation, different weights need to be assigned to the two TMs because of the nonlinear dependence of CARS to the pump and Stokes fields, as discussed above. The used coherent combination is therefore $TM_{coh} = \sqrt{2/3} \cdot TM(\omega_p) + \sqrt{1/3} \cdot TM(\omega_s)$.

To make the TM acquisition even faster and simpler, we finally developed a rather unconventional approach, based on the measurement of an 'incoherent' TM, called $TM_{incoh}$, measured by both pump and Stokes superimposed beams. For any incident wavefront, both pump and Stokes speckle patterns are likely to be uncorrelated and therefore in some areas, close to complementary where a bright speckle grain would correspond to either $\omega_p$ or $\omega_s$ maxima of intensity. The measurement of $TM_{incoh}$ is thus made of either $TM(\omega_p)$ or $TM(\omega_s)$ depending on the location, or an incoherent mixture of both. This scenario leads to enhancement factors of 150 and 130 for the $\omega_p$ and $\omega_s$ refoci respectively (Fig. 3(c)).

Ultimately, the different strategies described above lead all to a visible improvement of the balance between $\omega_p$ and $\omega_s$ enhancements, which is likely to be more favorable for CARS generation (Fig. 3(d)). The fact that their performances are comparable lies in the formation of the associated TMs, which is made to compensate the initial spectral decorrelation loss by a compromise for refocusing both $\omega_p$ and $\omega_s$ fields. Interestingly, the $TM_{interm}$ and $TM_{coh}$ strategies allow a fine tuning of the respective weight of pump and Stokes beams in the nonlinear coupling, which can be favorable for nonlinear frequency mixing processes in general.

We then used these refocused beams to generate a CARS signal in a layer of oil placed after the spinal cord slice (Fig. 3(e)). Oil is a model sample for evaluating coherent Raman generation at the $CH_2$ vibration bond present in lipids, corresponding to a wavenumber of 2845 cm$^{-1}$, which is reached at the anti-Stokes wavelengths of 645 nm. While no CARS signal was present before the beams refocus, nor with a wavefront refocusing $\omega_p$ with the use of $TM(\omega_p)$, we obtained a strong CARS generation following each of the three strategies described above, behind a 300 μm spinal cord sample. We ascertained the nature of the nonlinear generated signal by temporally delaying the $\omega_s$ beam with respect to the $\omega_p$ beam. A typical time response is displayed in Fig. 3(f). It follows a Gaussian like profile with width of about 420 fs, which corresponds to the expected time width from the nonlinear correlation occurring after propagation through the 300 μm thick sample. Note that while the level of CARS signal reached for all described strategies is similar, it can be further optimized using $TM_{interm}$ by an adequate choice of the intermediate wavelength used for the TM measurement (see Fig. S3 of Supplement 1).

At last we tested the capacities of the described strategies for point scanning over large field of views. Scans of the refocus were obtained for both pump and Stokes wavelengths, using the pre-knowledge of the TM applied for refocusing at all points in the sample exit plane. Fig. 4(a) shows measured intensity maps for both $\omega_p$ and $\omega_s$, as deduced from scanning a refocus in the image plane using a 200 nm-step (2 camera pixels in the image plane), through a 300 μm thick spinal cord tissue slice. For each TM strategy described above, correction for the heterogeneity of the reference speckles was ensured using a complementary speckle compensation method described in reference [24]. Fields of views of about 40x40 μm$^2$ are obtained, which surpasses the angular memory effect of the scattering medium [24] since the method is not based on spatial memory limitations. The Homogeneity of the obtained image is best for $TM_{interm}$ and slightly less performant for $TM_{coh}$, which is most probably due to the fact that a single correction for the speckle heterogeneities is performed for both wavelengths in this case. When using $TM_{incoh}$, the obtained intensity maps exhibit a poorer homogeneity, with complementary heterogeneities for $\omega_p$ and $\omega_s$. This is attributed to the nature of the construction of $TM_{incoh}$, which takes advantage of the propagation of either $\omega_p$ or $\omega_s$ to create a local bright spot, while methods used in $TM_{interm}$ and $TM_{coh}$ build-up a coherent reconstruction of a refocus for both wavelength simultaneously.

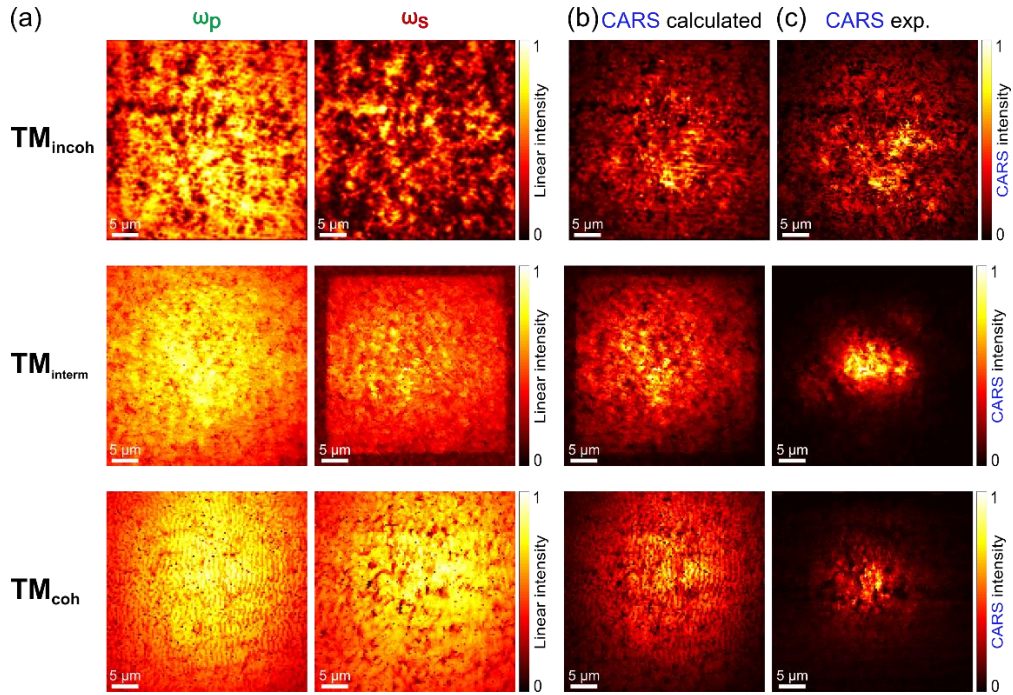

Fig. 4. Focus scanning through a 300 μm thick spinal cord tissue slice, placing an oil layer after the scattering medium. $\omega_p$ and $\omega_s$ are set at respectively 790 nm and 1019 nm, and was at 645nm (2845 cm$^{-1}$). The refocus is calculated each 400 nm (pixel size) in the image plane. (a) Linear scans showing obtained intensities for both $\omega_p$ and $\omega_s$ wavelengths, using refocusing strategies based on the TM$_{incoh}$ (top), TM$_{interm}$ (middle) and TM$_{coh}$ (bottom) (see text). (b) Calculated CARS image scans using the linear signals information shown in a). (c) Experimental CARS image.

At last, we demonstrate the capacity of our strategies to generate point scanning CARS images behind a scattering tissue. First, we simulated the maximum expected CARS scanned map using a computed CARS signal equal to $I_p^2 . I_s$. This supposes that both pump and Stokes fields are in phase in the image plane, therefore providing an upper limit for the expected CARS signal. This simulation leads to images of slightly reduced field of view as compared to the linear scans (Fig. 4(b)), which is due to the stringent dependence of CARS to high pump and Stokes intensities. We then generated experimental CARS maps in a sample made of a drop of oil, placed behind the spinal cord tissue slice, using the simultaneous scan of the $\omega_p$ and $\omega_s$ beams and filtering for the anti-Stokes wavelength. The produced signal is recorded pixel by pixel by a photomultiplier, which permits to reconstruct a CARS image (Fig. 4(c)). As expected, the obtained field of view is reduced as compared to linear images, reaching here about 25x25 μm$^2$, which is still superior to the expected angular memory effect range. The TM$_{interm}$ strategy is seen to produce a remarkably homogeneous CARS image as compared to TM$_{coh}$ and TM$_{incoh}$, as expected from its linear-intensities homogeneity. Overall the CARS image size is slightly smaller than the expected size, which is most probably due to the fact that at large distances from the optical axis, chromatic distortions make the spatial superposition of the $\omega_p$ and $\omega_s$ refoci more delicate [25].

## 3.     Discussion

Wavefront shaping for CARS has been previously studied in the frame of adaptive optics corrections, where the propagation regime was still in the ballistic regime where the $\omega_p$ and $\omega_s$ foci were deformed and not yet formed a speckle [5]. In particular, this condition required the optimization of a preliminary existing CARS signal, which is possible only for slight deformations of the optical wavefront. Producing a CARS signal in a scattering medium has been poorly explored, the main reason being that the manipulation of two wavelengths makes coherent control very delicate. In the present work, we show that a CARS signal can be recovered through a scattering medium even in conditions where both pump and Stokes beams are strongly affected by random beams propagation, e.g. in the form of uncorrelated speckles. By manipulating the wavefront of optical waves thanks to the measurement of the TM of the scattering media, it is possible to exploit such TM to recover refocus for both $\omega_p$ and $\omega_s$ frequencies, consequently producing a CARS signal.

Note that manipulating the wavefront independently for both wavelengths could be done using two SLM's, with strong disadvantages in terms of time required for the measurement, and set-up complexity. The approach proposed here allows an in-situ correction that does not require additional time overlap control. Another possibility could be to exploit the SLM voltage to phase relation to find an optimal voltage for each pixel that will accommodate both required wavelengths phase shifts [26]. A limitation of this approach is however the need for very large phase shifts to be provided by the SLM.

While this approach only addresses transmission geometries where the incident beams' manipulation is monitored from the other side of the sample, which is not yet reproducing the conditions of epi-microscopy, it permits to evaluate the first foundations of what the limits are of a workable wavefront shaping strategy for CARS, in terms of reachable intensities, scattering sample properties, and reachable wavenumbers. This study permits in particular to conclude that in conditions where CARS wavenumbers impose large wavelength distances of up to 250 nm, point scanning capabilities are still preserved through mice fixed spinal cord through thicknesses of up to 300 μm. This is a few times higher than what is reachable today in CARS microscopy, where depths of less than 80 μm are typically measurable in fixed samples (note that in fresh samples, larger depths are expected to be reached). This work demonstrates that coherently manipulating incident photons through thick media is possible for a very general range of nonlinear multi-frequency mixing generation. This also opens interesting routes towards the possibility to image CARS structures inside such samples, by the use of reflection strategies such as photo-acoustics or the reflection matrix.

## 4. Materials and methods

### A. Nonlinear wavefront shaping microscope

The CARS signals were generated by a fs-laser (Chameleon Ultra II, Coherent Inc.) tuned to 790 nm, with a pulse repetition rate of 80 MHz and 140 fs pulse width at the laser exit. It pumped an optical parametric oscillator (Compact OPO, APE / Coherent Inc.) that served as the Stokes beam, typically at 1019 nm, with a pulse width of 200 fs at the OPO exit. Due to dispersive elements of the microscope, the pulses widened temporally such that we estimate pulse widths of each beam to reach approximately ~300 fs at the microscope image plane (see Supplement 1). A motorized translation stage (M406-62S, Physik Instrumente) guaranteed the temporal overlap of pump and Stokes pulse trains. Pump and Stokes beam were combined by a 980LP dichroic beam splitter (Semrock) placed before the spatial light modulator (HSP256-0785, Boulder Nonlinear Systems). Its wavelength difference was set to 2845 cm$^{-1}$ which matches the vibrational resonance of $CH_2$ molecules. Accordingly, CARS photons were generated at a wavelength of 645 nm and detected with an analogue photomultiplier tube (R9110, Hamamatsu). The CARS signals where integrated over 10 ms per pixel. To prevent the excitation wavelengths from interfering with the CARS signal detection, a bandpass was designed of two short-pass (SP) filters (700+758, Semrock) and a 600 long-pass filter. The latter was added to hinder potential two-photon absorption signals (arising from the spinal cord tissue) from interfering with the CARS detection. To fit even large, millimetric biological tissues in the microscope, we chose to use a long working distance WD = 8.93 mm objective (Olympus 20x / 0.45 IR) for focusing light through the scattering medium. For collecting the scattered light with sufficient resolution and field-of-view, we chose a 40x / 0.6 objective (Olympus IR) that imaged the transmitted light via a tube lens onto a CMOS camera (BFLY-U3-23S6M-C, FLIR). The separation of excitation light (for measuring the transmission matrix) and CARS / SHG photons was mediated by a 757LP dichroic beam splitter. This allowed simultaneous monitoring of elastic scattered / linear photons as well as nonlinear scattered photons (in refocusing condition). CARS signals were generated with 50 mW pump and 80 mW Stokes beam power on the spinal cord tissues.

In the presence of glass coverslips that hold the sample and the scattering medium it is important to place the sample in the image plane of the excitation objective which can be done by looking at the reflection from the glass interfaces of the coverslips (next to the scattering medium where a reflection of the focus can be observed). To image the reflected light with maximum efficiency onto a CMOS camera (BFLY-U3-23S6M-C, FLIR), we placed a PBS and a removable quarter wave plate before the excitation objective. Once the image plane of the TM is found, the quarter-wave plate can be removed to obtain a linear polarization state in excitation and detection during the TM acquisition.

## B. Measuring the transmission matrix for refocusing

The spatial light modulator acts as the central device to shape the wavefront – for measuring the transmission matrix and refocusing the light behind the medium. Its optimal usage requires both pump and Stokes beams to overfill the active area of the SLM such that the Gaussian intensity profile of the laser beams appear flattened. This was achieved by expanding the beam after the laser exit. Thus, similar intensity is attributed to each pixel of the SLM. Because the generation of nonlinear signals requires the beam to be focused, the expanded beam on the SLM plane is conjugated onto the back aperture of the objective via a 4f-lens system. To obtain a pure linear polarization state parallel to the voltage controlled extraordinary axis of the liquid crystal of the SLM, we placed a polarizing beam splitter (PBS) in front of it.

The transmission matrix of the scattering medium was measured via self-referenced phase stepping interferometry. Here, the SLM is divided into two parts – the static reference and the active modulated part which is also used for refocusing after the TM has been measured. The static reference is co-propagating with the active modulated part, and is thus being scattered as well, which leaves the TM measurement spatially incomplete. We deploy complementary speckles to gain the capability to refocus light across the entire output field as presented in [24]. The TM was measured with typically 1024 Hadamard bases [8]. To get an accurate field measurement, less affected by noise, the phase at each basis got stepped through 0 to $2\pi$ in 10 steps. Combined with few milliseconds integration time for each phase step, a TM measurement for one wavelength typically took 10 min, and accordingly 30 min for the complementary TM. The TMs were measured in a linear vertical polarization state to prevent influencing the measurement by depolarized scattered light.

## C. Mouse spinal cord slices

All experimental and surgical protocols were performed following the guidelines established by the French Ministry of Agriculture (Animal Rights Division). The architecture and functioning rules of our animal house, as well as our experimental procedures have been approved by the "Direction Départementale des Services Vétérinaires" and the ethic committee (ID numbers #18555-2019011618384934 and A1305532 for animal house and research project, respectively). Mice were transcardially perfused with 4% paraformaldehyde. The spinal cords were dissected, post-fixed for 2 hours, and cut into 100 μm, 300 μm, 500 μm and 1000 μm transversal sections using a vibratome (Leica Microsysteme, Rueil Malmaison, France). Slices were then sandwiched between two 170 μm thick coverslips, embedded in Fluoromount (00-4958-02, Thermo Fisher). Coverslips were sealed with transparent nail polish.

**Funding.** This work has been supported by ANR-15-CE19-0018-01 (MyDeepCARS), ANR-10-INBS-04-01 (France-BioImaging), and the A*Midex Interdisciplinary project Neurophotonics.

**Acknowledgment.** The authors thank Franck Debarbieux (Institut des Neurosciences de la Timone, Marseille France) Siddharth Sivankutty (Institut Fresnel, Marseille) for precious help and discussions.

**Disclosure.** The authors declare no conflicts of interest.

See Supplement 1 for supporting content.

# Coherent anti-Stokes Raman scattering through thick biological tissues by single wavefront shaping

## Supplement 1

This document provides supplementary information to "Coherent anti-Stokes Raman scattering through thick biological tissues by single wavefront shaping". It contains an estimation of the chromatic error induced by the SLM, a detailed figure of the setup with a description of the measurement of the pulse width of refocused light and a derivation of the determination of the TM wavelength choice for optimal CARS signal generation.

## 1. Effect of the chromatic dependence of the SLM

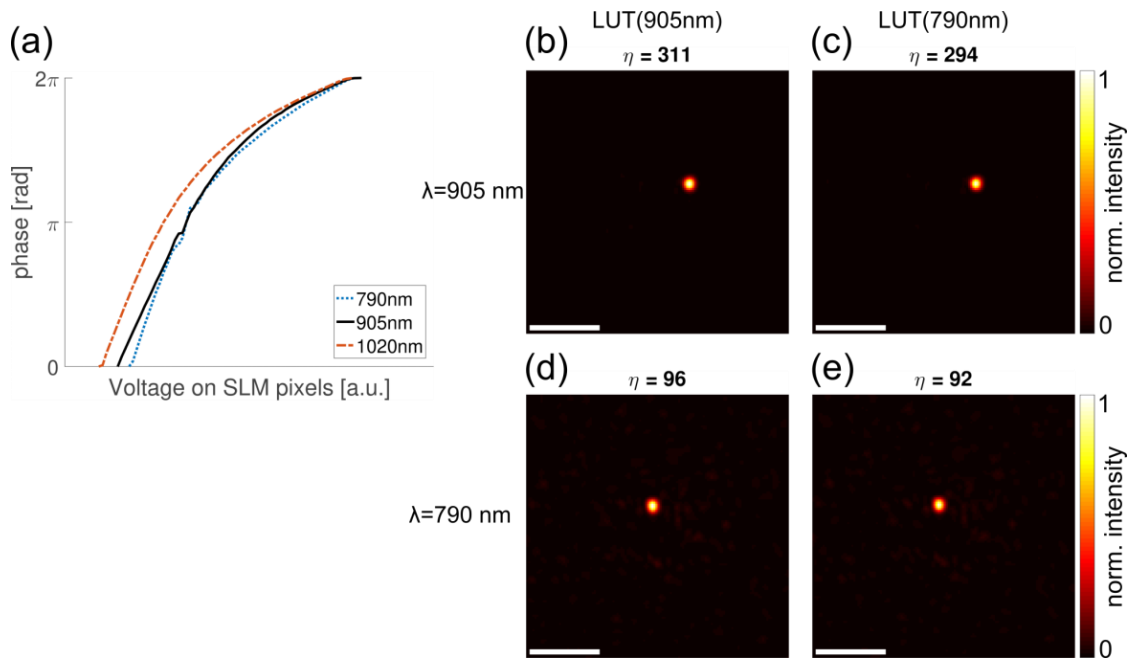

Fig. S5. (a) Look-up table (LUT) measured for three different wavelength. (b) to (e) Refocused beam through a 300 µm spinal cord tissue, using the TM measured at 905nm with the LUT for 905nm. (b) and (c) Refocus

with 905nm wavelength using a LUT of 905nm (b) and 790 nm (c). (d) and e) Refocus with 790nm wavelength using a LUT of 905nm (d) and 790nm (e). Scale bars are 5 μm.

In the experiments, one SLM was used to shape the wavefront for two distinct wavelengths of 790 nm and 1019 nm. To better distinguish the contribution of the spectral decorrelation of the scattering medium itself and the chromatic error that is introduced by the liquid crystal of the SLM itself, we characterized this error with the help of look up tables (LUT) for different wavelengths. A LUT establishes the relation between the voltage that is applied on a single pixel of the SLM and its phase response. This relationship is not linear as depicted in Fig. S1(a). Here, we measured the phase-voltage dependence for three different wavelengths - the pump wavelength 790nm, the Stokes wavelength of 1019nm and the mean of these two wavelengths 905nm. In the CARS experiments the LUT for 905nm to distribute the error induced by the SLM on both pump and Stokes beam equally. However, in Fig. S1(a) it is shown that the LUT for 905nm is closer to the one of 790nm than the one of 1019nm which indicates that the chromatic error is higher for 1020nm than for 790nm even when using the LUT of a mean wavelength.

To quantify the error, we perform a wavefront shaping experiment through a 300 μm spinal cord tissue where we measure the TM at a wavelength of 905nm using the LUT for 905nm. The refocus under this perfect wavelength match scenario shows an enhancement of 311 (see Fig. S1(b)). Using the very same phase mask for refocusing but applying the LUT of 790nm to the SLM leads to an enhancement factor of 294 (see Fig. S1(c)). A relative error of (311-296)/311=4.8% results from this. The error introduced here is therefore not coming from the scattering medium because the wavelength of refocusing remained unchanged, it stems from a chromatic mismatch in LUTs.

Under the same conditions, the wavelength was tuned to 790nm and the refocusing experiment was conducted with the LUT measured at 905nm and 790nm, depicted in Fig. S1(d) and (e) respectively. Due to the spectral decorrelation of the scattering medium, the enhancement factors drop to 96 and 92 respectively. The relative error resulting from the mismatch of LUTs is here (96-92)/96=4.2%. In conclusion, the change in enhancement due to chromatic error of the SLM is small compared to the medium spectral decorrelation.

## 2. Measurement of the pulse time width after refocusing through a scattering medium

To reconstruct the autocorrelation Second Harmonic Generation (SHG) traces, both beams are reflected on the SLM and refocused into a 200 μm-thick β-barium borate (BBO) crystal. In order to measure the second order interferometric autocorrelation as an auxiliary option to the existing CARS microscope, we designed it such that one can change between the two modes of measurement using a polarizing beam splitter (PBS, see Fig. S2), that either reflects the pump beam completely for CARS measurements (path A in dark green, see Fig. S2) or splits the pump beam into two paths (path B) for the measurement of the pump pulse width. Flip mirrors are used to guide light in either path A or B, and a removable half-wave plate and a PBS for recombining the split pump beam path. The half-wave plate after this PBS rotates both components by 45 degrees such that on passing through the first PBS, both the components are equally projected onto the SLM with a vertical linear polarization state. Note, half of the pump beam power is lost due to the selection of the vertical polarization component.

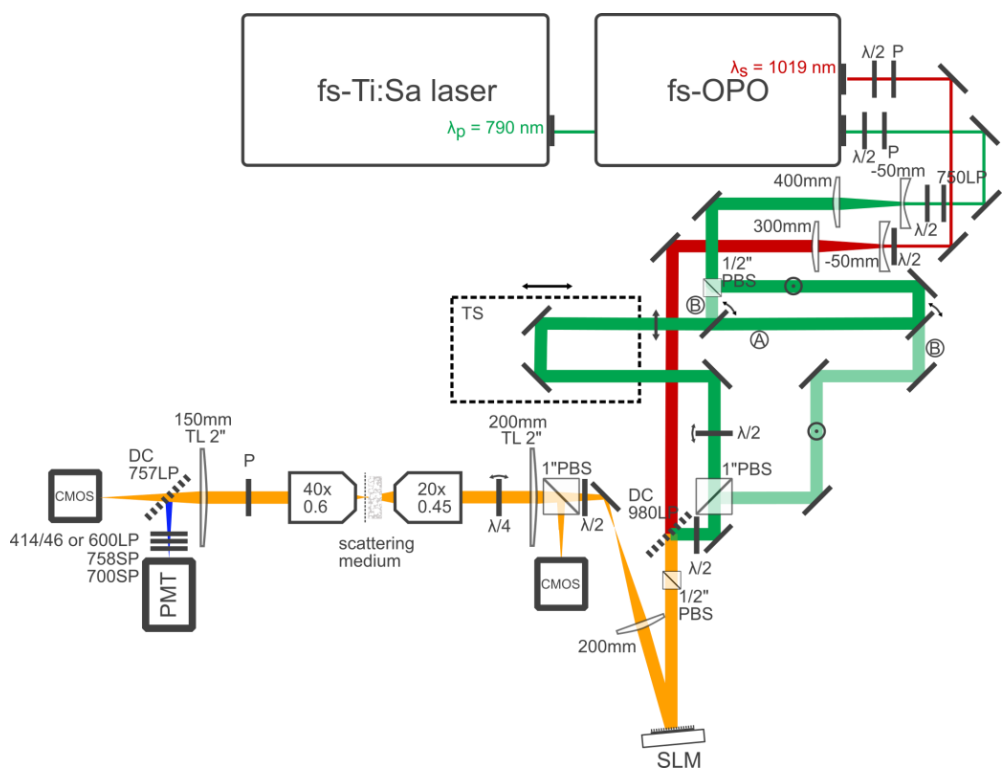

Fig. S6. Measurement of the pulse time width after refocusing through a scattering medium. A and B letters indicate the two beams from the fs pump laser, that are used for CARS measurement and autocorrelating SHG signals after refocusing, respectively. PBS: polarizing beam splitter. DC: dichroic mirror. TS: translation stage. SLM: spatial light modulator. LP/SP/BP: long/short/band pass filter. TL: tube lens. PMT: photomultiplier. sCMOS: camera.

The BBO crystal was attached to the coverslip of the scattering medium is used for the SHG readout signal. The second harmonic photons at 395nm wavelength were reflected onto the photomultiplier (PMT) by a dichroic mirror (DC 757 LP). A 414/46 BP, 758SP and 700SP filters on the detection path ensured that no residual excitation light reached the PMT. The TM was measured in a plane within the BBO crystal. Once a refocus was established by projecting the suitable phase mask onto the SLM, the second order interferometric autocorrelations were measured by translating the delay stage in steps of 50 nm, which is equivalent to 0.333 fs pulse delay in time. Note, that light propagates twice the distance of the delay stage because it is back reflected with a second mirror. The measured time interferogram is composed of a background plus an oscillating function. This function is averaged out by numerical averaging, which permits to access the intensity autocorrelation of the pulse width fitted by a Gaussian function, supposing here that the pulse is Gaussian.

## 3. Estimation of the required intensities balance for pump and Stokes beams, for optimal CARS generation

Having only one SLM at hand, and two distinct incident wavelengths, raises the question at which wavelength the TM should be measured to maximize the nonlinear signal. It is assumed that the incident laser power levels are already maximized, and we will provide in the following an additional method to increase the CARS signals by finding an optimal wavelength to measure (or construct) the TM of the scattering medium.

The derivation of the optimization procedure shall be visualized by the main parts' experimental conditions: Pump wavelength $\lambda_p$ = 790 nm, Stokes wavelength $\lambda_S$ = 1019 nm and a 300 μm spinal cord tissue as a scattering medium.

The consequence of measuring the TM at the pump wavelength is that the pump beam will have a high intensity focus whereas the Stokes beam will be poorly focused due to the limited spectral bandwidth of the scattering medium. In other words, the central wavelength at which the TM is measured determines the focus intensity balance for the pump and Stokes beam.

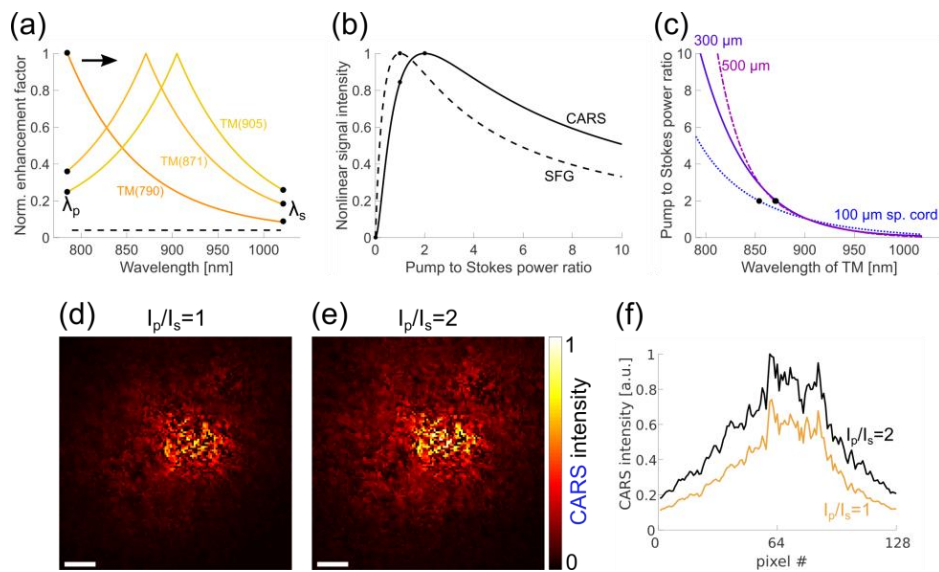

Fig. S7. (a) Spectral decorrelation curves of a 300 µm spinal cord tissue with different center wavelengths of the TM: 790 nm, 871 nm, 905 nm. The curves are normalized to their maximum value. (b) Optimal pump to Stokes power ratio for maximum CARS signal generation assuming a constant total intensity $I_p + I_s = 1$. A similar curve is plotted for Sum Frequency Generation (SFG, dashed line). (c) Pump to Stokes power ratio plotted as a function of the center wavelength of the TM. At a ratio of 2, maximum CARS signal generation is achieved (marked with black points for the different spinal cord thicknesses). (d) and (e) CARS point scanning in oil through 300 µm spinal cord with the coherent TM weighed equally (d) and in (e) with a twofold weight of the pump TM. (f) Mean over vertical image direction of (e). Scale bars are 5 µm.

This is illustrated in Fig. S3(a), where the expected enhancement factor at a given wavelength is plotted for different central wavelengths at which the TM is measured. These curves are deduced from the experimental fits obtained for a central wavelength of 790 nm (Fig. 2(b). The curves of 871 nm and 905 nm are shifted versions the 790 nm curve, assuming a symmetric decorrelation for longer and shorter wavelengths alike. All curves are moreover normalized. The black points mark the normalized enhancement factors expected for the pump and Stokes wavelengths. We read for instance from these curves, that at the mean wavelength of pump and Stokes of 905 nm, we would get an equal focus intensity for pump and Stokes. The dashed line marks the value experimentally obtained from the background speckle, at which level there is no more focus. Note that these curves are sample dependent (Fig. 2(b)).

However, the CARS signal $I_{CARS}$ is proportional to the product of the pump intensity $I_p$ squared and the Stokes intensity $I_s$: $I_{CARS} \propto I_p^2 \cdot I_s$, and thus the pump intensity has more weight. As a consequence a shift of the measured TM towards the pump wavelength will likely enhance the CARS signal. To calculate the optimal ratio of pump and Stokes intensity, we assume a varying $I_p/I_s$ ratio and $I_p + I_s = 1$. Inserted in the CARS intensity proportionality equation above, we plot the CARS intensity dependence as a solid black line in Fig. S3(b) with respect to $I_p/I_s$. We derive from it that the maximum CARS signal is achieved when giving the pump wavelength twice as much focus intensity as the Stokes wavelength. For comparison we plot the sum frequency generation (SFG) intensity as a function of the intensity ratio of pump and Stokes and find that both beams should be equally weighted to receive a maximum SFG signal. This is because SFG signals are a product of linear intensities of pump and Stokes beam: $I_{SFG} \propto I_p \cdot I_s$.

Since the slope of the spectral correlation curves is not linear, but follows an exponential decay, the optimum wavelength at which the TM is favorably measured for CARS needs to be calculated. We do this numerically by shifting the measured spectral correlation curve of Fig. S3(a) from the pump wavelength (790 nm) to the Stokes wavelength (1019 nm). That way we mimic a spectral decorrelation measurement at different center wavelength of the TM. For each of those curves an intensity ratio of pump and Stokes wavelengths can now be gathered and plotted as a function of the center wavelength (see Fig. S3(c)). The same procedure was performed not only for 300 μm spinal cord thickness but also for 100 μm and 500 μm. From this plot we deduce the wavelength at which the TM should be measured to get maximum CARS signals. This happens when the intensity ratio of pump and Stokes is 2 and thus, the optimal wavelengths are 855 nm, 871 nm and 868 nm for 100, 300 and 500 μm spinal cord thicknesses respectively. In this work, we use 866 nm as an intermediate wavelength for all measured samples.

The choice for optimal intermediate wavelength has been experimentally validated by composing a coherent TM from the pump and Stokes TMs and perform a CARS measurement through a 300 μm spinal cord tissue. Fig. S3(d)) and (e) depict the CARS intensities obtained from a coherent TM weighed equally, creating a power ratio of pump and Stokes of 1 or with a coherent TM where the pump TM is weighed doubled as compared to the Stokes TM which creates a power ratio of 2. To compare CARS signal strengths, we plot in Fig. S3(f) the cross sections of the row with the highest intensity pixel of Fig. S3(e). When weighing the coherent TM optimally, the maximum CARS signal is 1.25 times higher and the mean signal over the entire FoV has increased by roughly 1.5 times.